\documentclass{aa}

\usepackage{graphics,graphicx}
\usepackage{xcolor}
\usepackage{amsmath}
\usepackage{amssymb}
\usepackage{braket}
\usepackage{wrapfig}
\usepackage{caption}
\usepackage{subcaption}
\usepackage{natbib}
\newcommand{\kms}{km\,s$^{-1}$}

\bibpunct{(}{)}{;}{a}{}{,} 

\begin{document}
\title{The fountain of the luminous infrared galaxy Zw049.057 as traced by its OH megamaser}
\titlerunning{Nuclear dynamics of Zw049.057}
\author{Boy Lankhaar\inst{1,2} 
\and Susanne Aalto\inst{1} 
\and Clare Wethers\inst{1} 
\and Javier Moldon\inst{3,4} 
\and Rob Beswick\inst{4}
\and Mark Gorski\inst{1} 
\and Sabine K\"onig\inst{1} 
\and Chentao Yang \inst{1}
\and Jeff Mangum \inst{5}
\and John Gallagher \inst{6,7}
\and Francoise Combes \inst{8}
\and Dimitra Rigopoulou \inst{9,10}
\and Eduardo Gonz\'{a}lez-Alfonso \inst{11}
\and S\'ebastien Muller \inst{1}
\and Ismael Garcia-Bernete \inst{9}
\and Christian Henkel \inst{12}
\and Yuri Nishimura \inst{13}
\and Claudio Ricci \inst{14,15}
}
\institute{\inst{1}Department of Space, Earth and Environment, Chalmers University of Technology, Onsala Space Observatory, 439 92 Onsala, Sweden;
\email{boy.lankhaar@chalmers.se} \\
\inst{2}Leiden Observatory, Leiden University, Post Office Box 9513, 2300 RA Leiden, Netherlands \\
\inst{3} Instituto de Astrofísica de Andalucía, Glorieta de la Astronomía, s/n, E-18008 Granada, Spain \\
\inst{4} Jodrell Bank Centre for Astrophysics, Department of Physics and
Astronomy, The University of Manchester, Manchester M13 9PL,
UK \\
\inst{5}National Radio Astronomy Observatory, 520 Edgemont Road, Charlottesville, VA 22903, USA \\
\inst{6} Wisconsin IceCube Particle Astrophysics Center, Madison, WI
53703, USA \\
\inst{7} Department of Physics and Astronomy, Macalester College, 1600 Grand Ave, St. Paul, MN 55105 USA \\
\inst{8} Observatoire de Paris, LERMA, Collège de France, CNRS, PSL
University, Sorbonne University, Paris, France \\
\inst{9} Astrophysics, Department of Physics, University of Oxford, Keble Road, Oxford OX1 3RH, UK \\
\inst{10} School of Sciences, European University Cyprus, Diogenes Street, Engomi, 1516 Nicosia, Cyprus \\
\inst{11} Universidad de Alcalá, Departamento de Física y Matemáticas,
Campus Universitario, E-28871 Alcalá de Henares, Madrid, Spain \\
\inst{12} Max-Planck-Institut für Radioastronomie, Auf dem Hügel 69,
53121 Bonn, Germany \\
\inst{13} Department of Astronomy, The University of Tokyo, 7-3-1, Hongo,
Bunkyo, Tokyo 113-0033, Japan \\
\inst{14} Instituto de Estudios Astrof\'isicos, Facultad de Ingenier\'ia y Ciencias, Universidad Diego Portales, Av. Ej\'ercito Libertador 441, Santiago, Chile \\
\inst{15} Kavli Institute for Astronomy and Astrophysics, Peking University, Beijing 100871, China 
}
\date{}

\abstract
{High resolution ($0\farcs037-0\farcs13$ [$10-35$\,pc]) e-MERLIN ($\lambda6-18$ cm) and ($0\farcs024$ [$6.5$\,pc]) ALMA ($\lambda 1.1$ mm) observations have been used to image OH (hydroxyl) and H$_2$CO (formaldehyde) megamaser emission, and HCN $3 \to 2$ emission towards the nuclear ($<100$ pc) region of the luminous infrared galaxy Zw049.057. Zw049.057 hosts a compact obscured nucleus (CON), thus representing a class of galaxies that are often associated with inflow and outflow motions. Formaldehyde megamaser emission is detected towards the nuclear region, $\lesssim 30$ pc ($0\farcs1$), and traces a structure along the disk major axis. OH megamaser (OHM) emission is detected along the minor axis of the disk, $\sim 30$\,pc ($0\farcs1$) from the nucleus, where it exhibits a velocity gradient with extrema of $-20$ km\,s$^{-1}$ south-east (SE) of the disk and $-110$\,km\,s$^{-1}$ north-west (NW) of the disk. HCN $3 \to 2$ emission reveals extended emission, along the disk minor axis out to $\sim 60$\,pc ($0\farcs2$). Analysis of the minor axis HCN emission reveals high-velocity features, extending out to $600$\,km\,s$^{-1}$, redshifted on the SE side and blueshifted on the NW side. We propose that the high-velocity HCN emission traces a fast ($>250$\,km\,s$^{-1}$) and collimated outflow, that is enveloped by a wide-angle and slow ($\sim 50$\,km\,s$^{-1}$) outflow that is traced by the OHM emission. Analysis of the outflow kinematics suggests that the slow wide-angle outflow will not reach escape velocity and instead will fall back to the galaxy disk, evolving as a so-called fountain flow, while the fast collimated outflow traced by HCN emission will likely escape the nuclear region. We suggest that the absence of OHM emission in the nuclear region is due to high densities there. Even though OHMs associated with outflows are an exception to conventional OHM emission, we expect them to be common in CON sources that host both OHM and H$_2$CO megamasers.}


\maketitle

\section{Introduction}
The majority of the most luminous galaxies radiate the bulk of their energy in the far-infrared \citep[FIR, ][]{sanders:96}. According to their luminosity, these are either classified as luminous ($L_{\mathrm{IR}}>10^{11}$ $\mathrm{L}_{\odot}$) infrared galaxies (LIRGs), or ultra luminous ($L_{\mathrm{IR}}>10^{12}$ $\mathrm{L}_{\odot}$) infrared galaxies (ULIRGs). The extreme luminosities of (U)LIRGs are powered by an active galactic nucleus (AGN), an intense starburst (SB) phase, or both. (U)LIRGs emit the bulk of their energy in the FIR as their luminosity sources are embedded in a dusty phase that reprocesses radiation to longer wavelengths. While (U)LIRGs represent an evolutionary phase that many galaxies go through \citep[e.g.~][]{smail:97, hughes:98}, they themselves are rapidly evolving objects. 

A significant fraction of (U)LIRGs host compact ($r<100$\,pc), highly obscured ($N_{\mathrm{H}_2} \gtrsim 10^{25}$ cm$^{-2}$) and warm ($T \gtrsim 100$ K) nuclei \citep{aalto:15, falstad:19, falstad:21,garcia:22, donnan:23}. These compact obscured nuclei (CONs) are responsible for a large fraction of the total infrared luminosity of their host galaxies, despite their small sizes \citep{gonzalez:12, falstad:15, falstad:21}. Due to their opaque nature, it is difficult to discern the ultimate source of their luminosity. It has become increasingly clear that the CONs represent an evolutionary phase that is associated with both inflowing \citep{gonzalez:17, aalto:19, falstad:19} and outflowing \citep{barcos:18, falstad:18,gorski:23, wethers:24} gas motions. Yet, due to their opaqueness, signatures of radial gas motions are often only discernible at wavelengths $\gtrsim 1\ \mathrm{mm}$ \citep{falstad:19}. 

Radial gas motions towards galaxy nuclei are decidedly important to understand galaxy evolution. Inflowing gas motions drive nuclear activity and feed the nuclear star-forming gas. In turn, outflows transport the chemically enriched nuclear gas and dust away from the nuclear region, towards galaxy halos and the intergalactic medium \citep{veilleux:20, martin:05}, but also may contribute to nuclear growth by efficiently transporting angular momentum \citep{wada:12, konigl:94}. Galactic outflows may be driven by nuclear activity, stellar winds and supernovae from the SB phase, radiation pressure or cosmic-ray pressure \citep{veilleux:20}. Indeed, the association of CONs with radial gas motions indicates that these sources are rapidly evolving \citep{sakamoto:13}. Understanding these gas motions, and constraining their driving mechanism(s) is of pivotal importance to understand CONs, both as a subclass of (U)LIRGs, as well as their individual relations with their host galaxies. 

In this work, we study gas kinematics towards the galaxy Zw049.057 (IRAS\,15107+0724). Zw049.057 is a LIRG, with luminosity $L_{\mathrm{IR}} = 1.8 \times 10^{11}$ L$_{\odot}$, located ([RA, DEC] (J2000) = [15:13:13.10 +07:13:32.0]) at an estimated distance of $56$ Mpc \citep{sanders:03} that hosts a CON \citep{aalto:15, falstad:21}. We adopt a redshift of $z=0.01299$, which corresponds to a systemic velocity of $3897$ \kms\ \citep{katgert:98}. Throughout this paper the optical velocity convention is used with respect to the barycentric reference frame. It is an OHM galaxy \citep{baan:87, mcbride:13}, with evidence of inflows \citep{falstad:15} and outflow structures \citep{baan:87, falstad:18}, that are heavily obscured in the nuclear region \citep{falstad:19}. The major axis of the nuclear disk has a position angle $\mathrm{PA}_{\mathrm{maj}}=210^{\circ}$ \citep{falstad:18}. We present ALMA observations of the HCN $J=3 \to 2$ emission, that show signatures of a collimated outflowing structure, and e-MERLIN observations of OHM and H$_2$CO (formaldehyde) emission towards Zw049.057. We find evidence that the OHM emission is associated with an outflow structure, while population inversion, and thus the OHM emission, is suppressed in the nuclear disk region. 

OH megamasers are hosted by (U)LIRGs, where they are commonly taken to represent a compact and intense mode of star formation, that is likely triggered by tidal density enhancements due to merger events \citep{darling:07,darling:02}. The picture of \citet{darling:07} is indeed consistent with modeling efforts \citep{parra:05,lockett:08}, as well as previous high angular resolution imaging of the more luminous OHM using Very Long Baseline Interferometry (VLBI) instruments \citep{lonsdale:98,pihlstrom:01,richards:05}. This picture, though, is in tension with the association of OHM with an outflow structure; a feature of OHM that has been identified in an increasing number of sources \citep{baan:87, baan:89, gowardhan:18}. We explore this tension in detail by comparing resolved e-MERLIN imaging of the H$_2$CO megamaser of Zw049.057 with the resloved imaging of the OHM, and analyzing their (diss)association on the basis of previous excitation modeling \citep{vanderwalt:07, vanderwalt:22, lockett:08}, as well as the emerging picture of the inner dynamics of CON regions \citep{aalto:12,falstad:19,falstad:21,gorski:23}. 

This paper is structured as follows. In section 2, we present the observational set-ups that were used. In section 3, we describe the results of our observations. In section 4, we discuss our results, where we lay emphasis on the outflow dynamics, as well as the implication of our results for the class of OH megamasers. We present our conclusions in section 5.

\section{Observations}

\subsection{e-MERLIN observations}

Zw049.057 was observed with the e-MERLIN array between 7-14th January 2019 at L-band (project code  CY7222) and 7-8th January 2018 at C-band (CY6203). Observations included all e-MERLIN Telescopes, except the Lovell Telescope (6 telescopes in total). L-band observations were correlated into eight 64-MHz wide spectral windows spanning a total frequency range of 1245-1757\,MHz. At C-band observations encompassed 4488-5000\,MHz, correlated into 4 spectral windows. A total of 17.25 and 14.8~hours of observations were used at L-band and C-band, respectively.  

Correlators were set to the C-band 4829\,MHz H$_2$CO $J_{K_aK_c}$=1$_{10}$-1$_{11}$-transition and to the L-band 1667 and 1665\,MHz OH transitions. 
The flux calibrator was 3C286 and the baseline calibrator OQ208 for both sets of observations. Nearby gain calibrators 1516+0701 and 1513+0713 were used at L- and C-band respectively. 

Data were initially processed using NRAO's {\sc casa} \citep{CASATeam2022PASP} using the e-MERLIN pipeline\footnote{https://github.com/e-merlin/e-MERLIN\_CASA\_pipeline} \citep{moldon:21} which applies flagging, delay and bandpass calibration, and calculates phase and frequency-dependent amplitude gain corrections. Following initial calibration observing bands centered on the redshifted frequencies of the OH (1665 and 1667\,MHz rest frequency) and H$_2$CO (4830\,MHz rest frequency) lines were extracted and imaged to form individual spectral line cubes for further analysis. 

The synthesized beams are 0\farcs11$\times$0\farcs039 ($30 \times 11$ pc) for the H$_2$CO observations and 0\farcs26$\times$0\farcs13 ($70.6 \times 35.3$ pc) for the OH observations. The resulting data have a sensitivity of 0.2~mJy per beam in a 64~km\,s$^{-1}$ (1~MHz) channel width (H$_2$CO) and 0.77~mJy per beam in a 20~km\,s$^{-1}$ (0.11~MHz) channel width (OH).

\subsection{ALMA observations}
\label{s:obs_ALMA_B6}

Observations were carried out with 42 antennas in the array configuration C43-9/10, with projected baselines between 92~m and 13.8~km,  on Aug 28, 2021.  The on-source integration time was 34 minutes under relatively good atmospheric conditions (amount of precipitable water vapor $PWV \sim 0.5$\,mm). 

The bandpass response of the individual antennas was determined from the quasar $J$1550+0527. The quasar $J$1521+0420 was observed regularly for gain calibration. The absolute flux scale was calibrated using the quasar $J$1550+0527. The flux density for $J$1550+0527 was taken from the ALMA flux calibrator database.

The correlator was set up to cover four spectral windows in Band~6. The first two spectral windows were assigned to the upper side band, with bandwidths of 1.875~GHz and spectral resolution of 7.8~MHz, to cover the HCO$^+$ $J$=3$\to$2 and vibrationally excited HCN $J$=3$\to$2 $\nu_2$=1f lines (spectral window centered at sky frequency 264.3~GHz) and the HCN $J$=3$\to$2 line in the ground vibrational state (spectral window centered at 262.5~GHz). Two additional 2-GHz wide spectral windows were assigned to the lower side band, with a coarser spectral resolution of 31.25~MHz, to sample the continuum emission.

All of the ALMA Band~6 observations of Zw049.057 will be published in an upcoming paper by Wethers et al. (priv. com.). Here we present some results on HCN $J$=3 $\to$2.

The synthesized beam is 0\farcs027$\times$0\farcs024 ($7.33 \times 6.52$ pc), where Briggs weighting was used, adopting a robust parameter of 0.5. The resulting data have a sensitivity of 0.26~mJy per beam in a 20~km\,s$^{-1}$ (18~MHz) channel width.

\section{Results}
\subsection{L band continuum}
\begin{figure*}[p]
\centering
    \begin{subfigure}[b]{0.32\textwidth}
    \includegraphics[width=\textwidth]{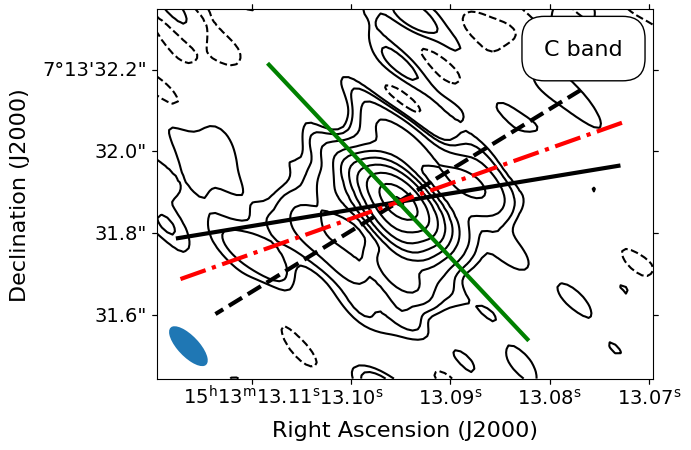}
    \caption{} 
    \label{fig:C_continuum}
    \end{subfigure}
\hfill
    \begin{subfigure}[b]{0.32\textwidth}
      \includegraphics[width=\textwidth]{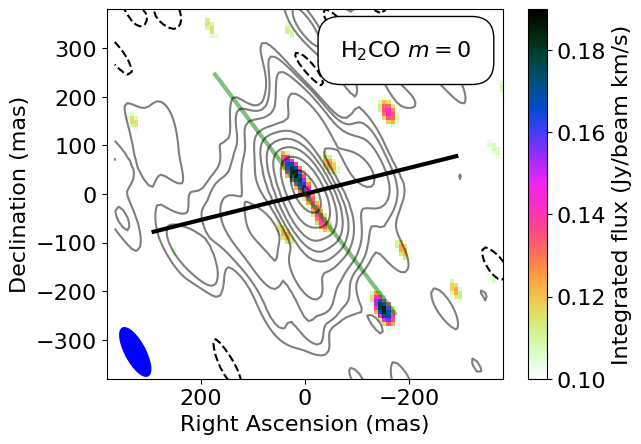}
      \caption{}
      \label{fig:H2CO_moment0}
    \end{subfigure}
\hfill
    \begin{subfigure}[b]{0.32\textwidth}
      \includegraphics[width=\textwidth]{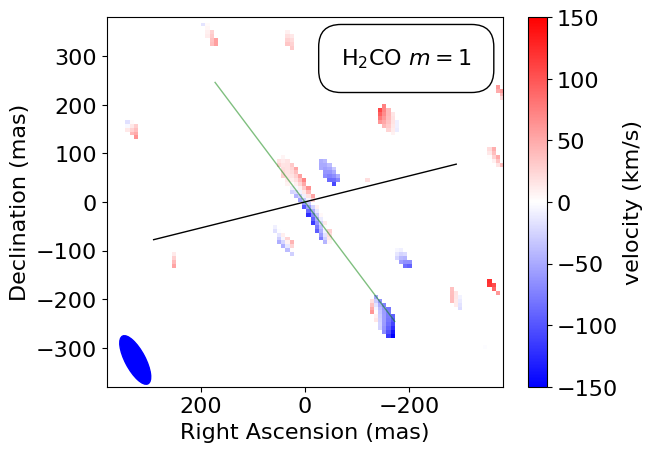}
      \caption{}
      \label{fig:H2CO_moment1}
    \end{subfigure}
\hfill
    \begin{subfigure}[b]{0.32\textwidth}
      \includegraphics[width=\textwidth]{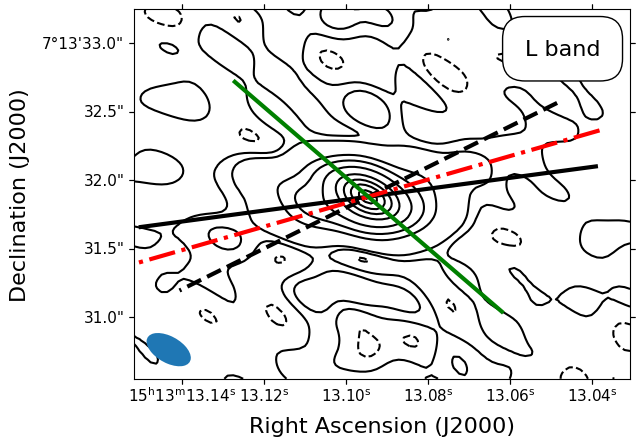}
      \caption{}
      \label{fig:L_continuum}
    \end{subfigure}
\hfill
    \begin{subfigure}[b]{0.32\textwidth}
      \includegraphics[width=\textwidth]{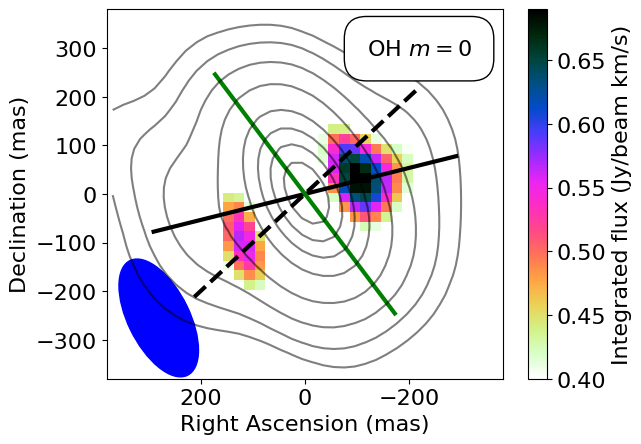}
      \caption{}
      \label{fig:OH_moment0}
    \end{subfigure}
\hfill    
    \begin{subfigure}[b]{0.32\textwidth}
      \includegraphics[width=\textwidth]{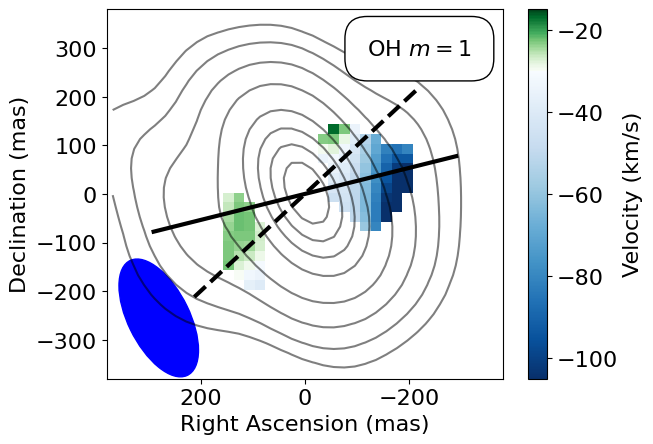}
      \caption{}
      \label{fig:OH_moment1}
    \end{subfigure}
\hfill
    \begin{subfigure}[b]{0.32\textwidth}
      \includegraphics[width=\textwidth]{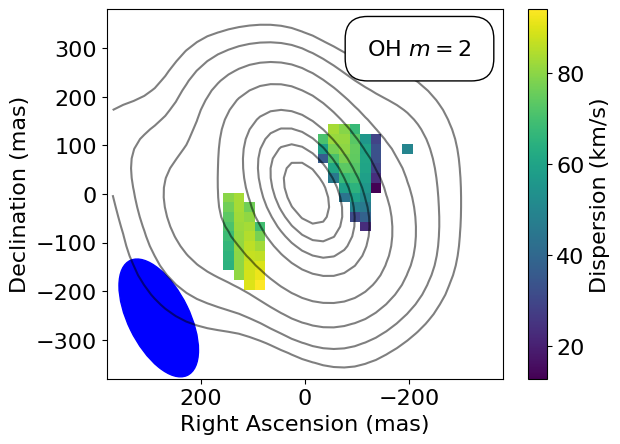}
      \caption{}
      \label{fig:OH_moment2}
    \end{subfigure}
\hfill
    \begin{subfigure}[b]{0.32\textwidth}
      \includegraphics[width=\textwidth]{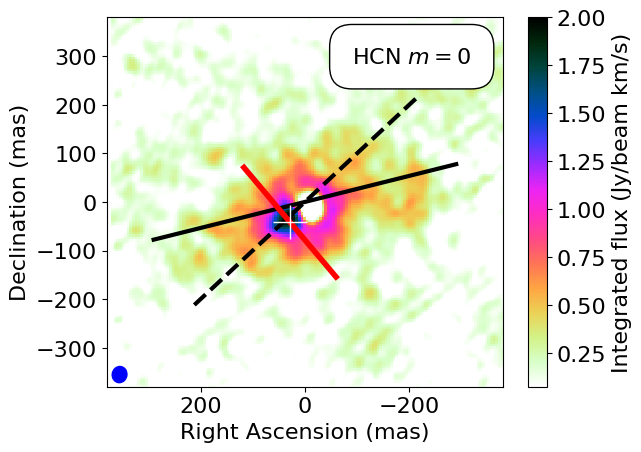}
      \caption{}
      \label{fig:HCN_0}
    \end{subfigure}
\hfill
    \begin{subfigure}[b]{0.32\textwidth}
      \includegraphics[width=\textwidth]{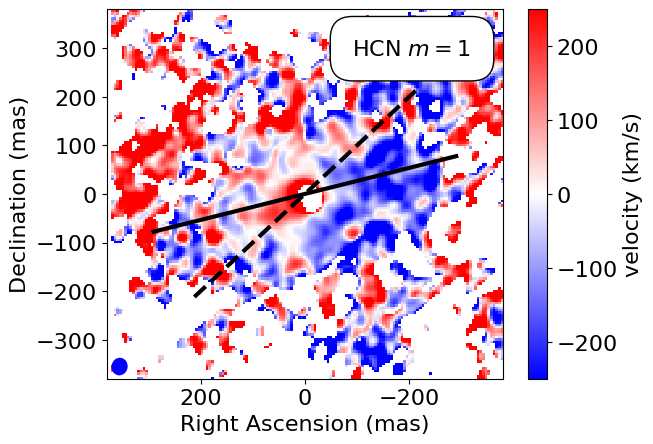}
      \caption{}
      \label{fig:HCN_1}
    \end{subfigure}
\hfill
    \begin{subfigure}[b]{0.32\textwidth}
      \includegraphics[width=\textwidth]{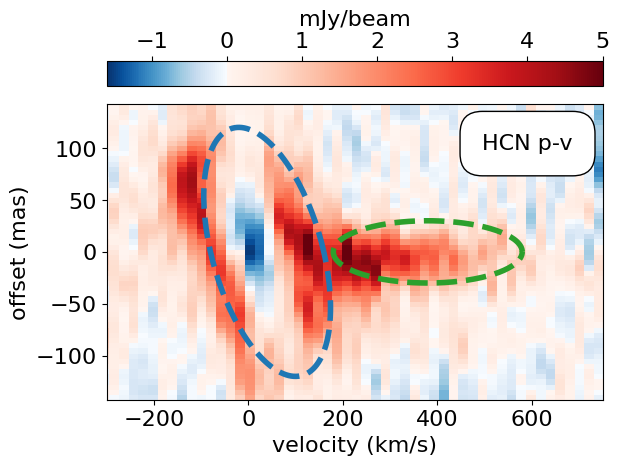}
      \caption{}
      \label{fig:HCN_pv}
    \end{subfigure}
\hfill
    \begin{subfigure}[b]{0.32\textwidth}
      \includegraphics[width=\textwidth]{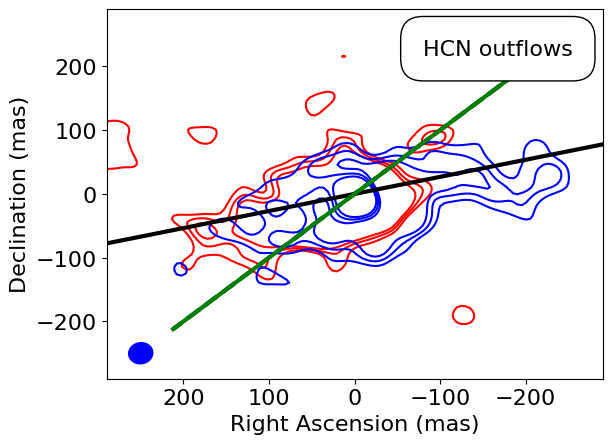}
      \caption{}
      \label{fig:HCN_redblue}      
    \end{subfigure}
    \begin{subfigure}[b]{0.32\textwidth}
      \includegraphics[width=\textwidth]{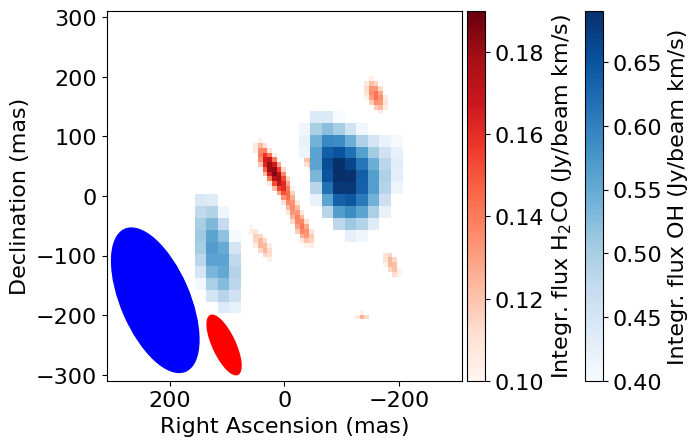}
      \caption{}
      \label{fig:OH_H2CO}
    \end{subfigure}
\caption{e-MERLIN and ALMA observations towards the nuclear region of Zw049.057. (a) e-MERLIN C band continuum contour map. Contour levels: [-1,1,3,30,60,120,240,360,$\cdots$]$\sigma$, with $\sigma=3.56 \times 10^{-5}$ Jy/beam. Dashed contour negative.
(b,c) H$_2$CO $4.83$ GHz emission moment 0 (b) and 1 (c) map overlaid (b) with C Band continuum emission contour map. (d) e-MERLIN L Band continuum emission contour map. Contour levels: [-1,1,3,6,12,24,48,96,384,$\cdots$]$\sigma$, with $\sigma=7.70 \times 10^{-6}$ Jy/beam.
(e,f,g) OH megamaser emission ($1.667$ GHz) moment maps (e, moment 0; f, moment 1; g, square root of the moment 2). Images are overlaid with L band continuum emission contours.
(h,i) Moment 0 (h) and moment 1 (i) maps of the HCN $J=3\to 2$ emission. In (h), negative emission yields (absorption) are put to zero. Red line and white mark in (h) indicate the slice and center of the p-v diagram shown in (j). (k) Integrated intensity of HCN high-velocity wings. Integration performed between velocities $\pm 250 - \pm 750$ \kms\ (blue/red). Contour levels: $[3,4,5,\cdots]\times 33$ mJy/beam \kms. (l) juxtaposition of the moment 0 maps of H$_2$CO (red) and OH (blue) emission.
Directions of the radio-jet (PA$_{\mathrm{jet}}=130^{\circ}$), CO outflow (PA$_{\mathrm{CO}}=105^{\circ}$), and disk major axis (PA$_{\mathrm{maj}}=30^{\circ}$) are indicated by the black dashed line, black solid line, and green solid line, respectively. Direction of the extended feature in the L-band continuum (PA$_{\mathrm{L}}=120^{\circ}$) is indicated with a red dash-dotted line. Axis systems of line emission maps are with respect to position of L band continuum maximum. Fitted beams (see text) are shown in the lower-left corner of the image.}
\end{figure*}

The L band continuum around $1.67$ GHz ($\lambda 18$ cm) was detected towards the nuclear region of Zw049.057; a contour map is given in Fig.~{\ref{fig:L_continuum}}. Continuum emission was found to exhibit extended features in the direction roughly perpendicular to the nuclear disk major axis, with position angle PA$_{\mathrm{C\ exten.}}=120^{\circ}$. The maximum flux density was observed as $21.5 \pm 0.4$ mJy\,beam$^{-1}$. We use the Rayleigh-Jeans law to obtain a corresponding brightness temperature of $3.25 \times 10^5$ K. The spatially integrated total flux density towards the nuclear region was $S_{\nu} = 41.4 \pm 4$ mJy, which can be compared to Arecibo single dish observations that retrieve an $18$ cm continuum flux of $40 \pm 5$ mJy \citep{baan:87}. Half of the total continuum emission emanates from an (angular) area of $0.05\ \mathrm{arcsec}^2$ ($60$ pc$^2$). We note that this area is very similar to the area of emittance of 33 GHz continuum radiation, $A_{33\ \mathrm{GHz}}=0.031$ arcsec$^2$, as found by \citet{song:22}. 

\subsection{OH megamaser (OHM)}
\begin{table*}[hbtp!]
\caption{Summary of the eMERLIN observations}
\label{tab:cont}
\begin{center}
\begin{tabular}{l c c c}
\hline\hline
                        & $I_{\nu,\mathrm{max}}$ (mJy beam$^{-1}$) $^{(a)}$   & $S_{\nu}$ (mJy) $^{(b)}$   & $\int dv S_{\nu}$ (Jy \kms) $^{(c)}$  \\ \hline
L band cont.            & $21.5 \pm 0.4$          &  $41.4 \pm 4$ &     \\
OH                      &  $6.50 \pm 0.77$     &     &  $2.0 \pm 0.4$            \\
C band cont.           &  $12.0 \pm 0.1$         &  $25.6 \pm 2$  &  \\
H$_2$CO                 &  $1.1 \pm 0.2$      &     &  $0.48 \pm 0.04$            \\
\hline                                                      
\hline
\end{tabular}
\end{center}
\textbf{Notes:} $^{(a)}$maximal flux density, $^{(b)}$spatially integrated flux density, $^{(c)}$spatially and spectrally integrated flux density. 
\end{table*}

OHM emission was detected towards the nuclear region of Zw049.057 in both the $1.667$\,GHz and the $1.665$\,GHz transitions. For the main maser transition at $1.667$\,GHz, we retrieved an integrated total flux of $2.0 \pm 0.4$ Jy\,km\,s$^{-1}$, while for the satellite maser transition at $1.665$\,GHz, we retrieved $1.0 \pm 0.2$ Jy\,km\,s$^{-1}$. The total integrated flux of the $1.667$\,GHz maser compares well to \citet{baan:87}, who find 2.25\,Jy\,km\,s$^{-1}$ and \citet{mcbride:13}, who find 2.45\,Jy\,km\,s$^{-1}$\,\footnote{We note \citet{mcbride:13} report an integrated flux $4.9$\,Jy\,km\,s$^{-1}$. However, as explained in \citet{mcbride:13} they use a different definition of the total flux density resulting in a factor two increase in the reported flux density compared to commonly used definitions including those used in this paper.

For a deeper discussion surrounding this topic, we refer to \citet{robishaw:21}.}. The integrated total flux corresponds to an OH luminosity of $L_{\mathrm{OH}}=10.1\ L_{\odot}$. Figure \ref{fig:OH_moment0} shows the moment 0 map of the $1.667$ GHz OHM emission. OHM emission is observed near the minor axis, with maximal emission appearing about 
0\farcs2 ($\sim 57$ pc) from the $18$ cm continuum peak. The emission to the west of the continuum peak appears stronger, 
compared to the emission to the east of the continuum peak. Emission towards the nucleus, as well as along the major axis, is undetected. The moment 1 map of the $1.667$ GHz OHM emission, given in Fig.~(\ref{fig:OH_moment1}) shows a clear velocity gradient, from $-20$ \kms\ in the east, to $-100$ \kms\ in the west (corresponding to about $5$ spectral channels). Within the NW OHM feature, a velocity gradient from the North to the South is observed, but we note that this feature occurs on a scale smaller than the beam size. The velocity dispersion, taken as the square root of the moment 2 map of the $1.667$ GHz OHM emission, given in Fig.~(\ref{fig:OH_moment2}), appears constant around $70$ km\,s$^{-1}$ throughout the OHM region. 

Imaging of the 1.665 GHz transition is given in Fig.~(\ref{fig:OH_1665}). It is readily apparent that the NW feature of the OHM main line is found also for the weaker 1.665 GHz maser line; the moment 0 image shows a feature at a similar position, while the moment 1 map shows a similar velocity field compared to the main maser transition at 1.667 GHz. We do not detect the SE feature of the OHM emission. We ascribe this to sensitivity effects, as for the main maser line, this feature was also considerably weaker.

\subsection{C band continuum}
The C band continuum around $4.8$ GHz ($\lambda 6.2$ cm) was detected towards the nuclear region of Zw049.057. In Fig.~(\ref{fig:C_continuum}) we display a contour map of the C band continuum. As for the L band continuum, an extended structure appears present along the minor axis (P.A.$\sim 120^{\circ}$). The maximum spectral flux density we observed as $12.0 \pm 0.1$ mJy/beam, corresponding to a brightness temperature of $1.68 \times 10^5$ K. We retrieve an integrated total flux density towards the nuclear region was observed as $S_{\nu} = 25.6 \pm 2$ mJy, which compares well to the MERLIN observations of \citet{baan:17} that retrieved $29.72$ mJy, as well as the Arecibo observations of \citet{araya:04} that retrieved $30.8$ mJy. Half of the total e-MERLIN C band continuum emission emanates from an (angular) area of $0.008\ \mathrm{arcsec}^2$ ($20$ pc). 

\subsection{H$_2$CO megamaser}
H$_2$CO megamaser emission was detected towards the nucleus of Zw049.057 in the ($J_{K_a,K_c} \to J'_{K_a',K_c'}$) $1_{1,0} \to 1_{1,1}$ transition at $4.83$ GHz. Figs.~(\ref{fig:H2CO_moment0}) and (\ref{fig:H2CO_moment1}) present the moment 0 and 1 maps of the detected emission, while a zoom-in of these figures can be found in Fig.~(\ref{fig:H2CO_zoom}). Maximal emission was found at 1.08 mJy beam$^{-1}$, corresponding to $1.4\times 10^4$ K. We retrieved an integrated total flux of $0.48 \pm 0.04$ Jy\,km\,s$^{-1}$, where \citet{baan:17} finds $0.308$ Jy\,km\,s$^{-1}$ using MERLIN and \citet{araya:04} find $0.283$ Jy\,km\,s$^{-1}$ using Arecibo. From the integrated total flux we compute the corresponding maser luminosity $7.3\pm 0.6\ L_{\odot}$, which is large enough for it to fulfill the threshold to be considered megamaser emission\footnote{Typical galactic H$_2$CO masers have luminosities $\lesssim 10^{-6}\ L_{\odot}$.} Figure \ref{fig:H2CO_moment0} shows the moment 0 map of the $4.83$ GHz H$_2$CO emission. H$_2$CO emission is observed concentrated along the major axis (P.A.$\sim 210^o$, \citet{falstad:18}), with maximal emission appearing about 
0\farcs1 ($\sim 29$ pc) NE of the $6.2$ cm continuum peak. Emission to the north of the $6.2$ cm continuum peak appears $\sim 25\%$ stronger than the emission to the south of the C band continuum peak. The moment 1 map of the emission, shown in Fig.~(\ref{fig:H2CO_moment1}) shows a velocity gradient, red to blue, from the North to the South.

\subsection{HCN minor axis emission}
The HCN $J=3 \to 2$ transition was detected toward the nucleus of Zw049.057. An image of the integrated intensity is given in Fig.~(\ref{fig:HCN_0}). In the inner $50$ mas ($14$ pc) towards the nucleus, HCN is observed in absorption, while it is observed in emission farther out from the nucleus. Emission extends farther along the minor axis (P.A.$\sim 120^o$, \citet{falstad:18}), compared to the major axis (P.A.$\sim 30^o$, \citet{falstad:18}). 

We analyzed the velocity structure of the minor axis emission using a position-velocity diagram (p-v diagram) of a strong feature along the minor axis. The position-slice is perpendicular to the minor axis, and indicated in Fig.~(\ref{fig:HCN_0}), where the zero-offset position is marked with a blue cross. The p-v diagram is plotted in Fig.~(\ref{fig:HCN_pv}). In the diagram, we observe a feature that presents a correlated position-velocity structure between $-150$ to $200$\,km\,s$^{-1}$, and a feature that is associated with high-velocity emission $>200$\,km\,s$^{-1}$, extending to $\sim 600$\,km\,s$^{-1}$ where we observe emission concentrated around the central position. We indicated the high-velocity feature by a green marking. Interestingly, the emission of the feature between $-150$ to $200$\,km\,s$^{-1}$ follows the profile of a skewed and shifted ellipse in PV-space. The ellipse, that was fitted by eye, is drawn in dashed blue in Fig.~(\ref{fig:HCN_pv}), and is $120$\,km\,s$^{-1}$ wide and $120$ mas high, shifted by $40$\,km\,s$^{-1}$, and skewed by a velocity gradient of $0.5$ \,km\,s$^{-1}$\,mas$^{-1}$. 

Finally, we present in Fig.~(\ref{fig:HCN_redblue}) an integrated intensity image of the high-velocity ($\pm 250$\,km\,s$^{-1}$ $-$ $\pm 750$\,km\,s$^{-1}$) emission. Both the blueshifted and the redshifted emission show an emission structure that is concentrated along the minor axis, consistent with the earlier discussed p-v diagram. We retrieved an integrated total flux of $12.7$ Jy\,km\,s$^{-1}$ for the redshifted high-velocity emission and $12.5$ Jy\,km\,s$^{-1}$ blueshifted high-velocity emission, but note that the integrated total flux of the blueshifted emission may be an overestimation due to contamination of the $4_{1,3} \to 3_{1,2}$ methanimine line (rest frequency $266.270024$ GHz) at $432$ \kms\ with respect to the HCN $3\to 2$ rest frequency. 

\section{Discussion}
\subsection{HCN outflow}
The HCN emission along the minor axis traces two distinct structures, as can be readily ascertained from Fig.~(\ref{fig:HCN_pv}). We first discuss the low velocity feature, indicated in blue in \ref{fig:HCN_pv}). From the p-v diagram, emission could be extracted that appeared as a shifted, skewed ellipse (width: $120$ \kms, height: $120$ mas, shift: $40$ \kms, skew: $0.5$ \kms\, mas$^{-1}$) in p-v space. We propose that this feature traces a rotating, wide-angle, and hollow, outflowing structure. The skew corresponds to a velocity gradient ($5.7 \times 10^{-14} \ \mathrm{s}^{-1}=1.8$ \kms\, pc$^{-1}$) that is due to the outflow rotation. The width and shift of the ellipse are due to emission from the front- and back side of the outflow, with a velocity directed radially from (width, $\frac{1}{2} \times 120$ \kms$=60$ \kms) and along (shift, $40$ \kms), the outflow symmetry axis \citep[see Appendix B of][for modeling of a transverse pv cut for an expanding wind annulus]{tabone:20}. We considered the possibility that the blue side of the elliptical structure represents the rotating nuclear disk, but ruled out this scenario, as the elliptical structure is present also in minor axis p-v slices $>200$ mas, beyond the HCN disk radius of $\sim 150$ mas.


Both Figs.~(\ref{fig:HCN_pv}) and (\ref{fig:HCN_redblue}) clearly indicate that the high-velocity emission emerges from a linear and confined structure. From its spatial and velocity structure, we interpret the high-velocity HCN emission along the minor axis as a fast collimated outflow. For the position angle of the collimated outflow we find $\mathrm{PA_{HCN}}=105^o$. Interestingly, this is the same as the position angle that \citet{falstad:18} find for the outflow traced by CO $J=6\to 5$ emission. \citet{falstad:18} find that the CO $J=6\to 5$ emission extends to $400$ \kms, less than the $\sim 600$ \kms\ we find, but this discrepancy may be due to different excitation conditions or sensitivity effects. We propose the CO $J=6\to 5$ emission, observed by \citet{falstad:18} and the high-velocity HCN emission are tracing the same structure. 

To estimate the mass of the fast collimated outflow, we extracted its luminosity and used the $L$(HCN) to $M$(dense) conversion relation of \citet{gao:04}. We note that the conversion relation of \citet{gao:04} was derived for the HCN $1-0$ line; to convert our HCN $3 \to 2$ luminosity to an HCN $1-0$ luminosity, we assumed a HCN $\frac{J=3\to2}{J=1\to0}$ temperature ratio of $0.3$ \citep{krips:08}. Accounting only for the red outflow ($12.7$ Jy\,km\,s$^{-1}$, $250-500$ \kms), this  procedure yields a collimated outflow mass of $M_{\mathrm{HCN}}=6.2\times 10^7\ M_{\odot}$. We note that the conversion relation is derived for gravitationally bound gas, so the mass will be overestimated if the dense gas is in non self-gravitating clumps. \citet{aalto:15} studied the impact of turbulence on the conversion factor and found that it should be scaled down by about a factor of five for non self-gravitating gas in the outflow. Because of these uncertainties, we adopt $M_{\mathrm{HCN}}=1-6\times 10^7\ M_{\odot}$ in order to compute the outflow kinematics from our mass estimates. Furthermore, we put conservative estimates of $\sim 250$ \kms\ and $30$ pc for the outflow velocity and size. Combined, we compute the mass-loss, momentum, momentum flux, energy and energy flux of the outflow as: $\dot{M}_{\mathrm{HCN}}=85.2-511\ \mathrm{M_{\odot}/yr}$, $p_{\mathrm{HCN}} = 2.5-15 \times 10^{9} \ \mathrm{M_{\odot} \,km /s}$, $\dot{p}_{\mathrm{HCN}} = 1.3-6.5\times 10^{12} \ \mathrm{L_{\odot}}/c$, $E_{\mathrm{HCN}} = 0.8-3.9 \times 10^{55} \ \mathrm{ergs}$ and $\dot{E}_{\mathrm{HCN}} = 0.5-2.7 \times 10^{9} \ \mathrm{L_{\odot}}$. We note that we have not included any energy contribution from gas dispersion in our estimates for the outflow energy. In terms of the IR luminosity of Zw049.057, we may put the moment flux and energy flux, $\dot{p}_{\mathrm{HCN}}/(L_{\mathrm{IR}}/c) \approx 7-34$ and $\dot{E}_{\mathrm{HCN}}/L_{\mathrm{IR}} \approx 0.003-0.015$. These are strikingly high yields for the momentum and energy fluxes. Compared to the sample of ULIRG outflows that \citet{gonzalez:17} analyzed, where outflow momentum and energy fluxes in the range of $2-5\ L_{\mathrm{IR}}/c$ and $0.1-0.3\%\ L_{\mathrm{IR}}$ were derived, both the energy and momentum fluxes of the outflow traced by HCN are at the higher end, or exceeding, the yields of the ULIRG outflows. We discuss this point in more detail in section \ref{sec:out_launch}.



\subsection{OHM outflow} 
The OHM spatial and velocity structures, showing emission along the minor axis with a velocity gradient of $\sim 100$\,km\,s$^{-1}$, red in the SE towards blue in the NW, are consistent with an outflowing structure. We rule out that the OHM is excited in the disk, as in this scenario maximal emission is expected towards the disk major axis \citep{pihlstrom:01, parra:05}, opposite to what we observe. Indeed, with velocity dispersion of the order of the disk rotation \citep{falstad:18}, the longest, velocity-coherent, columns of OH are expected towards the disk major axis, while Fig.~(\ref{fig:C_continuum}) also shows significant seeding radiation to emerge therefrom. Yet no maser emission is observed from the disk, leading us to suggest that population inversion is suppressed there. While we cannot rule out that rotation contributes to the velocity profile of the OHM features, the observed velocity gradient cannot be explained by rotation alone. Finally, the evidence for outflowing structures emerging from the nuclear region in the HCN emission, including a structure that has similar kinematical properties as the features traced by the OHM, is additional circumstantial evidence that OHM is tracing an outflow structure. Thus, we interpret the OHM emission as emerging from an outflow structure.

We proceed to make kinematical estimates of the OHM outflow. In order to estimate the OHM outflow mass, we use the result that OHM operate around OH column densities of $N_{\mathrm{OH}}\sim 10^{17} \mathrm{cm}^{-2}$ \citep{lockett:08}. We define the OHM emission region as the one where the spectrally integrated intensity exceeds $0.3\ \mathrm{Jy/beam}$ \kms, which has an angular area of $0.12 \ \mathrm{arcsec}^2$. Using an abundance of $x_{\mathrm{OH}} = 2 \times 10^{-6}$ \citep{gonzalez:12, gonzalez:17} relative to H$_2$, and an average particle mass of $\mu = 2.3 \ \mathrm{m_H}$, we compute the OHM outflow mass as $M_{\mathrm{OHM}} = 9 \times 10^6\ \mathrm{M_{\odot}}$. Adopting an outflow velocity of 50~km\,s$^{-1}$ and a size of 30~pc, the mass expulsion rate of the outflow is $\dot{M}_{\mathrm{OHM}} = 15\ \mathrm{M_{\odot} / yr}$. With these estimates, we compute the momentum, momentum flux, energy and energy flux of the outflow as: $p_{\mathrm{OHM}} = 8.8 \times 10^{46}$~g\,cm\,s$^{-1}$, $\dot{p}_{\mathrm{OHM}} = 3.7 \times 10^{10} \ \mathrm{L_{\odot}}/c$, $E_{\mathrm{OHM}} = 2.187 \times 10^{53}$~ergs and $\dot{E}_{\mathrm{OHM}} = 3.1 \times 10^6 \ \mathrm{L_{\odot}}$. In terms of the IR luminosity of Zw049.057, we may put the moment flux and energy flux, $\dot{p}_{\mathrm{OHM}}/(L_{\mathrm{IR}}/c) = 0.2$ and $\dot{E}_{\mathrm{OHM}}/L_{\mathrm{IR}} = 2 \times 10^{-6}$.

The OHM outflow velocity (50~\kms) is much less than the minimal velocity required to escape the inner region ($\sqrt{2}v_{\mathrm{circ}}\approx 200$ \kms,\ see, e.g.~\citet{martin:05}). Indeed, as discussed in section \ref{sec:out_launch}, the OHM outflow will likely fail to escape the inner region and will be obscured when observed in the IR range of the spectrum. Instead, the outflow will likely fall back onto the gas close to the nuclear region, causing a significant stirring of the nuclear gas \citep{aalto:20,wada:12}. Not only will the nuclear gas be endowed an appreciable amount of kinetic energy in the form of turbulence, but the gas stirring will also impact the accretion dynamics significantly, as turbulence is directly related to the effective viscosity of the gas \citep{shakura:76,pringle:81}. CON regions are consistently associated with nuclear gas velocity widths in excess of $100$ \kms. If this is dispersion, it would imply that the turbulent kinetic energy of the gas far exceeds the thermal energy: $E_{\mathrm{kin}}/E_{\mathrm{therm}}>10^4$. It may therefore well be possible, at least locally, that the conversion of turbulent energy through shocks may be a significant source of luminosity. We however note that the flows may still be largely laminar and that the high velocity widths are caused by line of sight effects of a layered structure of in- and outflows.

Spectroscopic IR observations, performed with Herschel, at 23 H$_2$O and 12 OH transitions, have not been able to definitively indicate the outflow we observe in the OHM emission \citep{falstad:15}. This is partially due to resolution and the relatively low outflow velocity of the OH gas. But, also, and importantly, we find clear evidence that the OHM emission is excited only in the outflowing gas. Thus revealing a map of the outflowing gas that is relatively uncontaminated by the disk emission. While we present the first resolved interferometric maps of OHM in association with an outflow, spectral features that were offset in velocity gave early indications in single dish observations of Zw049.057 \citep{baan:87} and other OHM galaxies \citep{baan:89}. Notably, through spectral analysis, \citet{gowardhan:18} were able to find striking evidence of the association of OHM with outflowing gas towards two ULIRGs, by combining unresolved OHM observations with interferometric imaging of CO and CN emission. In section \ref{sec:OH_exc} we discuss OHM tracing outflow structures in relation to their excitation conditions and often-observed association with H$_2$CO megamasers. 

\subsection{Outflow launching} 
\label{sec:out_launch}
The kinematic estimates we presented above may help us to constrain the launching mechanism of the high-velocity collimated outflow and the slow wide-angle outflow. First, we consider the scenario of a radiatively driven outflow. Radiation pressure is a momentum conserving mechanism to launch the outflows. We compare the total momentum flux of the outflow components and relate it to the momentum flux that emanates from the nuclear region. For the OHM outflow, we estimated earlier $\dot{p}_{\mathrm{OHM}}/(L_{\mathrm{IR}}/c) = 0.04-0.2$, while for the collimated HCN outflow, we estimated $\dot{p}_{\mathrm{HCN}}/(L_{\mathrm{IR}}/c) \approx 7-34$. Thus, without a significant momentum boost, it is difficult to explain the collimated HCN outflow with a momentum conserving outflow launching mechanism. 

It is possible for the wide-angle outflow to be radiatively driven. We explore the hypothesis that the slow wide-angle outflow traced by OHM is radiatively launched. First, the OHM clumps have significant column densities, that ensure continuum optical depths of order unity, even for long wavelengths. Indeed, with optical depths $\tau_{\mathrm{OHM}}\sim 1$ at $10 \, \mathrm{\mu m}$ (corresponding to the maximal specific intensity of thermal radiation of $\sim 300$\,K), it is a possibility that the OHM clumps are lifted from the CON surface by the thermalized IR radiation field. At the instance of our observations, the OHM outflow has attained a (projected) velocity of $v_{\mathrm{OHM}} \approx 50$ \kms. It is interesting to speculate about the progression of the outflow launching, assuming it is radiatively driven. If the outflow is to be accelerated, we require the outward force on an OHM clump, that is due to the radiation field, to exceed the inward force due to gravity. We follow \citet{heckman:15} and define the outward force due to the radiation field as $F_{\mathrm{out}} = \dot{p}_{\mathrm{rad}} A_c / 4\pi r^2$, where $\dot{p}_{\mathrm{rad}}$ is the radiation momentum flux, $A_c$ is the cloud area and $r$ is the distance from the nucleus. The inward force due to gravity we put at $F_{\mathrm{in}}=M_{\mathrm{OHM}} v_{\mathrm{cir}}^2 / r$, where $M_{\mathrm{OHM}}$ is the OHM clump mass and $v_{\mathrm{cir}}$ is its circular velocity. The critical  momentum flux to drive an OHM clump outwards can be obtained by equating the inward and outward forces and is $\dot{p}_{\mathrm{crit}} = 2.1 \times 10^{11} L_{\odot}/c \left(\frac{r}{30 \ \mathrm{pc}}\right) \left( \frac{N_{\mathrm{OHM}}}{3 \times 10^{22} \ \mathrm{cm}^{-2}}\right)\left(\frac{v_{\mathrm{circ}}}{140 \ \mathrm{km\ s}^{-1} } \right)^2$, adopting typical values for the OHM clumps that occur in the outflow as we observe it. Interestingly, the critical momentum flux is similar to $L_{\mathrm{IR}}/c$. This indicates that if the radiation is the driving agent, the OHM outflow is not accelerating. In fact, after numerical integration of the above equations, adopting the cited typical values for the OHM clumps that occur in the outflow as we observe it, we found that after $\sim 4\times 10^5$ years, the outflowing OHM clumps will acquire a negative velocity, indicating that they will evolve to fall back onto the galactic disk as a fountain flow \citep{wada:12}.
The collimated HCN outflow requires a driving mechanism beyond radiation, due to the high momentum requirements. First, the possibility has been explored, that the molecular outflow of Zw049.057 as traced by CO $J=6\to 5$ emission, which is likely the same outflow as traced by the high-velocity HCN emission, is entrained by a radio jet \citep{falstad:18}. Indeed, \citet{falstad:18} find that the radio continuum around 5.3 GHz shows an extended feature along the minor axis, and while this possible radio jet is not aligned with either the HCN or the CO outflow found by \citet{falstad:18}, it may still have powered them and deflected. Based on the jet power estimate, $\dot{E}_{\mathrm{jet}}=4.7 \times 10^{9} \ L_{\odot} \approx \dot{E}_{\mathrm{HCN}}$ \citep{falstad:18,baan:06,birzan:08}, we do not rule out the (tentatively detected) radio-jet as a driver of the dense collimated outflow, but we do note that our estimates for the energy flux are likely lower limits, as we have not included the velocity dispersion in the outflow energy estimate, and we use a projected velocity for the outflow velocity. Alternatively, the molecular outflow that we observe can be entrained by an energy driven AGN wind \citep{silk:98, aalto:12, aalto:15b}. If this is the case, we expect the collimated outflow to be associated with an ionized and hot wind. In addition, in order for the molecular material not to be disrupted by the entrainment, rapid radiative cooling and/or stabilizing magnetic fields are required \citep{cooper:09, mccourt:15, leaman:19}. 

The projected outflow velocity of the dense collimated outflow traced by HCN was put at $>200$ \kms, with maximal projected velocities approaching $650$ \kms. These outflow velocities are comfortably larger than $\sqrt{2}v_{\mathrm{circ}}\approx 200$ \kms, suggesting that this outflow will escape the nuclear region. Outflow velocities in excess of $3v_{\mathrm{circ}}\approx 420$ \kms\ are required to escape the galaxy \citep{martin:05}, which thus appears to be the fate of a significant fraction of this outflow. 

It has been suggested that the morphology and kinematics of CON regions resemble the protostellar environment \citep{aalto:20,gorski:23}. In protostellar regions, outflows are thought to be launched magnetocentrifugally \citep{blandford:82,pelletier:92,wardle:93,spruit:96,tabone:20}, and magnetically collimated. The outflow takes the appearance of a fast collimated jet-like structure, emerging from the inner launching regions, while towards the outer launching regions, a wide(r)-angle and lower velocity outflow emerges \citep{spruit:96, tabone:20}. Such an appearance is both consistent with the dynamics and chemistry we find for our outflows, but we note that this mechanism operates most effectively when gas flows are laminar and critically depends on the presence of an ordered magnetic field. 

Magnetic fields in cool molecular gas may either be probed through dust polarization observations \citep{andersson:15}, or through spectral line polarization, either through the Zeeman effect \citep{crutcher:93,lankhaar:23} or through the Goldreich-Kylafis effect \citep{goldreich:81,lankhaar:20a}. Unfortunately, continuum emission is heavily affected through self-absorption and scattering effects in buried regions \citep{gonzalez:19}, even at (sub)millimeter wavelengths. With OHM tracing the outflow, observation of its circular polarization would lead to constraints on the magnetic field in these regions. Such extragalactic Zeeman observations of OHM have been performed previously \citep{robishaw:08}, on a sample of OHM sources exhibiting strong emission. For the Zw049.057 OHM, with its modest intensity yields, Zeeman observations will require very deep integration, which would be beyond feasibility for the current radio telescopes, but may become possible with the next generation of radio observatories such as the SKA. Alternatively, it may be possible to attain magnetic field information towards the nuclear region through linear polarization observations of vibrationally excited HCN emission, that is known to be luminous \citep{falstad:21} and expected to be significantly polarized due to its radiative excitation \citep{lankhaar:20a,lankhaar:22a}.  

\subsection{Inner dynamics of Zw049.057}
\begin{figure}[tbh]
 \includegraphics[width=9cm, trim = 5cm 0cm 5cm 0cm, clip]{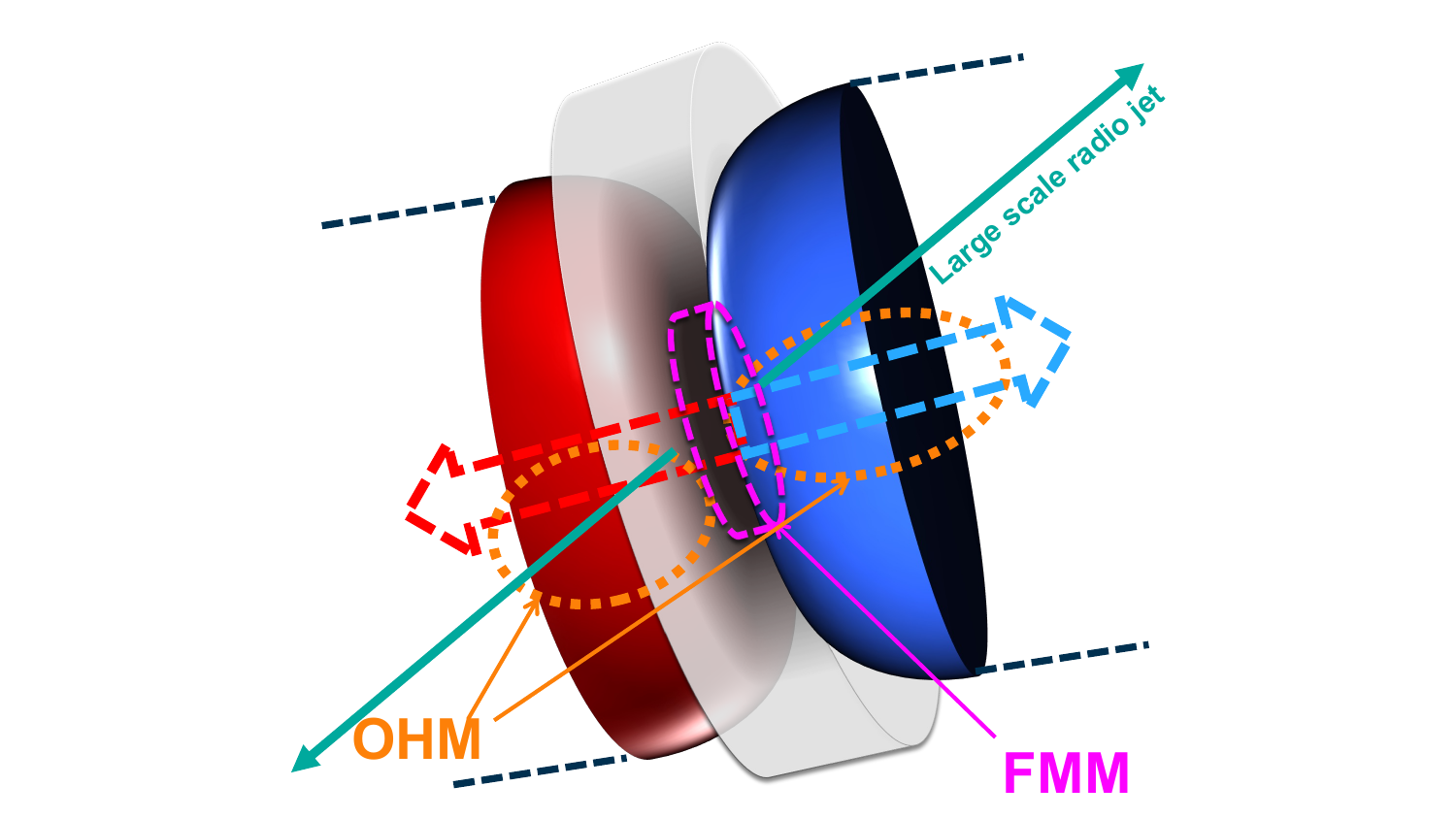}
  \caption{Cartoon of nuclear system of Zw49.057. The collimated dense outflow that is traced by the high-velocity HCN emission is indicated with dashed red/blue arrows. Rotating wide-angle wind as traced by the lower velocity HCN emission is indicated by the hollow bi-spheres. Dotted lines are added to the ends of the hollow bi-cones to indicate their extension further out. The outflowing OH megamaser emission is excited towards the regions indicated by dotted orange ellipses. H$_2$CO megamaser emission (FMM) is excited in the inner regions of the nuclear disk, indicated by dotted magenta. We have furthermore drawn the extended nuclear disk in semitransparent gray-scale, as well as the large-scale radio jet, that is differently inclined from the outflow system traced by HCN and OH. }
  \label{fig:cartoon_zw}
\end{figure}

Based on the spatial and velocity signatures of the H$_2$CO megamaser, OH megamaser and HCN emission, as well as constraints from excitation modeling that we discuss in section \ref{sec:OH_exc}, we propose the following model of the dense gas surrounding the nucleus of Zw049.057. Fig.~(\ref{fig:cartoon_zw}) presents a cartoon of the dense gas surrounding the nucleus of Zw049.057.  In this model the H$_2$CO megamaser is tracing the nuclear disk\footnote{Our imaging of H$_2$CO maser emission towards Zw049.057 agrees very well with the imaging performed by \citet{baan:17} using MERLIN. Yet, \citet{baan:17} find for other H$_2$CO megamaser galaxies other configurations of H$_2$CO maser emission, that does not seem to be tracing exclusively the nuclear disk. In the absence of interferometric data to a larger sample of H$_2$CO megamaser galaxies, we do not intend to make the general claim that H$_2$CO megmaser emission traces the nuclear disk of the host galaxy.}. The OH megamaser is not excited towards the nuclear disk, but rather, we propose, in a wide-angle and slow moving outflow structure. 
The OHM outflow envelopes a denser, faster, and collimated outflow that is traced by high-velocity HCN emission. This is likely the same outflow that \citet{falstad:18} detected in CO $J=6\to5$ emission at lower resolution. As discussed in section \ref{sec:out_launch}, we estimate that the dense collimated outflow has likely attained sufficient velocity to escape, at least, the nuclear region. In contrast, the wide-angle outflow traced by the OH megamaser does not possess sufficient velocity to escape the nuclear region. 
Instead, we expect the wide-angle outflow to be drawn back to the nuclear disk after $\sim 1$~Myr to form a so-called fountain flow \citep{wada:12}.

\subsection{OH megamasers}
\subsubsection{OH megamaser excitation}
\label{sec:OH_exc}
It is interesting to compare our observations of the $1.667$ GHz OHM to existing observations of OH towards Zw049.057. \citet{falstad:18} observed the OH $4.660$ GHz, $4.766$ GHz and $4.751$ GHz hyperfine transitions in absorption towards the nucleus of Zw049.057. Of these transitions, the $4.751$ GHz is the main hyperfine transition, while the transitions at $4.660$ GHz and $4.766$ GHz are satellite hyperfine transitions.  The relative intensities of these three hyperfine transitions are (4.660:4.751:4.766) GHz = (1:2:1) in the optically thin limit, under local thermodynamic equilibrium \citep[LTE; ][]{destombes:77}. The relative peak absorption of the lines follow the ratio expected under LTE.  However, different line profiles were found among the transitions, which was interpreted as the result of OH gas that is excited under varying conditions, comprising the signal \citep{falstad:18}. Furthermore, \citet{falstad:18} found that the spatially and kinematically resolved emission from the $4.751$\,GHz transition, indicate that the dominant part of the emission emerges from a rotating structure along the major axis, while they found tentative evidence for a feature at $\sim 200$ \kms\ to the northwest of the continuum peak. \citet{mcbride:13} observed the satellite transitions associated with the $1.67$ GHz OHM at $1.612$ and $1.712$ GHz using the Arecibo telescope. From the conjugate line profiles of the OHM satellite lines, they concluded that two separate inversion mechanisms exist in Zw049.057.

Earlier, we showed that the OHM appears in an outflowing structure along the galactic minor axis. The OH outflow appears to envelop a collimated and dense outflow traced by high-velocity HCN emission.
While OH is observed towards the nuclear disk in its $6$ cm and far-IR lines \citep{falstad:15, falstad:18}, we detect it there neither as a maser nor in absorption in the $1.667$ and $1.665$ GHz transitions. This, while H$_2$CO is observed as a megamaser in association with the nuclear disk. The combination of these observations confirms earlier conjectures, that distinct excitation mechanisms/regions exist for the OH emission towards Zw049.057 \citep{mcbride:13, falstad:18}.

The presence and absence of OHM and H$_2$CO maser emission allows us to constrain the excitation conditions in the outflow and nuclear region of Zw049.057. It is well established that OH megamasers appear in association with FIR radiation around 53 $\mu \mathrm{m}$ absorbed through the OH $^{2}\Pi_{\frac{1}{2}}-$$^{2}\Pi_{\frac{3}{2}}$ $J=\frac{3}{2} - \frac{3}{2}$ transitions, that causes population inversion in the transitions around $1.67$ GHz through radiative pumping \citep{lockett:08}. Population inversion is achieved in turbulent clumps ($\Delta v$$\sim 20$ kms$^{-1}$) of sizes $\lesssim$ 1 pc, number densities $\lesssim 10^5$ cm$^{-3}$ and specific OH column densities of the order $10^{16} \ \mathrm{cm}^{-2} \mathrm{km}^{-1} \mathrm{s}$. For such clumps, maser optical depths $\tau_{1.667\ \mathrm{GHz}} \sim -1$ are expected \citep{lockett:08}. A bright (non-thermal) background radiation field around $18$ cm wavelengths is required for OH megamasers to be apparent, while multiple maser clumps along the line-of-sight may boost the amplification of this background radiation significantly \citep{parra:05}. Detailed modeling of the excitation of H$_2$CO masers finds that the $4.8$ GHz H$_2$CO maser is excited towards denser regions $\gtrsim 10^{5} \ \mathrm{cm}^{-3}$, where the maser pumping is collisional \citep{vanderwalt:22}. Maser pumping can operate in environments with a prominent FIR radiation field, provided the gas temperature exceeds the radiation temperature \citep{vanderwalt:22}. A rather restrictive constraint that \citet{vanderwalt:22} put to the operation of H$_2$CO masers, is that it requires H$_2$CO to be present in high abundance ($x_{\mathrm{H_2CO}} \sim 10^{-5}$). Such H$_2$CO abundances are $3$ to $4$ orders of magnitude higher than what is typically found towards Galactic star forming regions, hot cores or dark clouds \citep[see, e.g.,][]{mangum:93, tang:18, kirsanova:21}. We note, that the abundance constraint was formulated for H$_2$CO masers excited towards high-mass star-forming regions. For H$_2$CO megamasers, velocity-coherence length scales, $\ell$, can be two orders of magnitude higher, while line-widths, $\Delta v$, are one order of magnitude higher. Thus, taking into consideration that (large-velocity gradient) radiative transfer problems depend primarily on the specific column density, $x_{\mathrm{H_2CO}} n_{\mathrm{H}_2} \ell / \Delta v$, the abundance requirements are relaxed by about an order of magnitude. Still, taking this into consideration, H$_2$CO abundance requirements seem too high, and we suggest dedicated modeling of H$_2$CO megamaser sources is required to address this issue.

We suggest that the absence of OH emission or absorption around $1.67$ GHz in the nuclear disk is the consequence of the high densities in these regions, that effectively suppress the maser mechanism. Instead, because of the outflow expansion, OH clumps in the outflow structure are more tenuous, allowing thus for efficient maser pumping. In addition, for the OHM to operate, one requires efficient FIR pumping. The FIR radiation in the OH outflow may either emerge from the CON region, or from the collimated denser part of the outflow that is traced by high-velocity HCN emission, and which the OH outflow envelops. While the CON region subtends a large solid angle as seen from the OH outflow, absorption may diminish its prominence. On the other hand, considering the lifetime ($\sim 10^5$ years) of the collimated outflow, we require a mechanism such as star formation in this region to heat up the dust phase, to account for the emergence of FIR radiation here from. 

\subsubsection{The class of OH megamasers and CONs}
A synthesis of the OHM surveys by \citet{baan:89b} and \citet{darling:02} and the host galaxy IR, CO and HCN luminosities, indicate an association between OHM and high molecular gas densities and fractions, as well as an unusual IR-CO luminosity relation \citep{darling:07}. On the basis of this, it is suggested that OHM are associated with a compact and intense mode of star formation, that is likely triggered by tidal density enhancements due to merger events \citep{darling:07}. This picture is consistent with modeling efforts \citep{parra:05,lockett:08}, and are not in tension with previous high resolution imaging of the OHM using VLBI instruments \citep{lonsdale:98,pihlstrom:01,richards:05}, that were performed on, primarily, the more luminous OHM. However, our high-resolution observations of the OHM emission that show the association with an outflowing structure towards the LIRG Zw049.057, require a more subtle picture of the class of OHM. 

We pointed out earlier that Zw049.057 is a CON, a subclass of (U)LIRGs that host exceptionally dense nuclear regions and are associated with both inflow and outflow motions. While only a small fraction ($\lesssim 10\%$) of LIRGs, and up to a third of ULIRG galaxies host OHM \citep{darling:02}, almost all LIRG and ULIRG CON galaxies that have been searched for OHM emission, indeed host OHM. At first, this is not surprising, as OHM are associated with high density gas \citep{darling:07}, and show prominent absorption in OH around $53 \ \mathrm{\mu m}$, which is a requisite for OH maser pumping \citep{gonzalez:12, gonzalez:15}. Yet, of the known H$_2$CO megamasers, 5/6 are hosted by CONs, of which 1 of them is a ULIRGs (Arp\,220) and the 4 others are LIRGs (IC\,860, UGC\,5101, IRAS\,17578–0014, Zw049.057), where each of them is additionally associated with OH maser emission. The association of the OHM with H$_2$CO megamaser appears problematic, as only very limited conditions allow for the simultaneous excitation of both H$_2$CO and OH megamasers. Our high-resolution interferometric observations show that the issue of the apparent association between OHM and H$_2$CO megamaser emission can be resolved, as they emerge from different regions; outflow (OHM) and nuclear disk (H$_2$CO megamaser) structures. We speculate that a similar arrangement may be present for the other LIRGs that host OH and H$_2$CO megamaser.

\subsection{OH megamasers at high redshift}

With this expected surge in OHM detections in distant galaxies, it is natural to discuss the implications of the current study. From the interferometric mapping of the OHM towards Zw049.057, our main finding is its association with an outflow, and the suppression of emission towards the nuclear disk. The outflow traced by OHM is expanding at a low velocity ($\sim 100$ km/s), making it difficult to detect spectroscopically. This is in contrast to other studies that have associated OHM with outflows \citep{baan:89, gowardhan:18, glowacki:22}, that found OHM emission at large ($200-1000$ km/s) offsets from the systemic velocity. The luminosity of the OHM of Zw049.057 ($10.1\pm 2.0\ L_{\odot}$) is on the lower end of the range of OHM luminosities. This feature can be ascribed to (i) Zw049.057 being a LIRG, with a naturally weaker radio-continuum that is to be amplified by OHM, (ii) the slight offset of the OHM from the continuum maximum, and (iii) the relatively small region that is conducive to population inversion. In particular, points (ii) and (iii) are a direct consequence of OHM tracing an evolved outflowing structure.

A useful feature of the untargeted HI surveys, is their broadband scanning. Thus, it is likely that for an OHM galaxy, not only the main OHM lines at $1.67$ GHz will be detected, but also satellite transitions at 1612 and 1720 MHz. A comprehensive analysis of the different transitions can reveal properties of the (pumping) conditions towards the OHM galaxy \citep{baan:87}. Such an analysis has already been productively applied by \citet{hess:21} to an OHM found in the HI survey AWES (Apertif Wide-area Extragalactic imaging Survey). For Zw049.057, analysis of the OHM lines and their satellites, revealed the presence of different pumping conditions for the OHM; a feature which we can confirm through our interferometric observations. Multi-line analysis of the OHM lines and their satellites will reveal important information on the pumping conditions, but to formulate rigorous relations that reveal the OHM association with an outflow, more interferometric studies such as this one are necessary. To direct additional interferometric studies, we have pointed to H$_2$CO megamaser emission as a marker for OHM galaxies with OHM emission tracing outflow structures, similar to Zw049.057.

\section{Conclusions}
We present high-angular resolution interferometric observations of the kinematics of the nuclear region of Zw049.057. Using e-MERLIN, we observed the H$_2$CO and OH megamasers, and using ALMA, we observed HCN $3\to2$ emission. The H$_2$CO megamaser traces the nuclear disk, while the OH megamaser traces a slow and wide-angle outflow. The slow and wide-angle OHM outflow in turn appears to envelop a fast and collimated outflow that is traced by high-velocity HCN emission. 

Momentum estimates indicate that the OH megamaser outflow may be radiatively launched, while the fast and collimated molecular outflow traced by HCN requires an additional momentum boost. Such a momentum boost may be achieved if the outflow is entrained by an energy-conserving AGN wind, or through a magneto-hydrodynamical launching mechanism. From the low velocity of the OHM outflow and an analysis of the driving forces, we conclude that the wide-angle OHM will likely not reach escape velocity, and instead evolve as a fountain flow. 

We observed the H$_2$CO and OH megamaser to be associated with different physical structures. The excitation of OH and H$_2$CO megamasers require different conditions, that are almost mutually exclusive, yet of the known H$_2$CO megamasers, all of them have been found towards OH megamaser galaxies. Additionally, five of the six known H$_2$CO megamasers are found towards CONs. Since the prime characteristic of CONs is dense nuclear gas, and they are commonly associated outflow systems, we speculate that a similar arrangement can be expected for the OH-H$_2$CO (hydroxyl-formaldehyde) megamaser galaxies.

Our high-angular resolution interferometric observations have revealed multiple phases of outflowing gas emanating from the inner ($\lesssim 10$ pc ), and obscured ($N_{\mathrm{H}_2} \sim 10^{25}$ cm$^{-2}$), nuclear regions of Zw049.057. While the ultimate driving mechanisms of the outflowing gas phases remain ambiguous, it is clear that they impact the gas dynamics and mass and angular momentum budget of the nuclear region greatly. 

\begin{acknowledgement}
This paper makes use of the following ALMA data: ADS/JAO.ALMA\#2019.1.01612.S. ALMA is a partnership of ESO (representing its member states), NSF (USA) and NINS (Japan), together with NRC (Canada), MOST and ASIAA (Taiwan), and KASI (Republic of Korea), in cooperation with the Republic of Chile. The Joint ALMA Observatory is operated by ESO, AUI/NRAO and NAOJ. The National Radio Astronomy Observatory is a facility of the National Science Foundation operated under cooperative agreement by Associated Universities, Inc. BL acknowledges support for this work from the Swedish Research Council (VR) under grant number 2021-00339. S.A., C.W., M.G.~and C.Y gratefully acknowledges support from an ERC AdvancedGrant 789410.
CR acknowledges support from Fondecyt Regular grant 1230345 and ANID BASAL project FB210003.
J.M. acknowledges financial support from the grant CEX2021-001131-S funded by MCIU/AEI/ 10.13039/501100011033 from the grant PID2021-123930OB-C21 funded by MCIU/AEI/ 10.13039/501100011033 and by ``ERDF A way of making Europe''. We thank the anonymous refere for constructive feedback that improved the quality of the paper.
\end{acknowledgement}

\bibliographystyle{aa}
\bibliography{lib.bib}

\begin{appendix}
\section{Additional figures}
\begin{figure*}[ht!]
\centering
    \begin{subfigure}[b]{0.45\textwidth}
    \includegraphics[width=\textwidth]{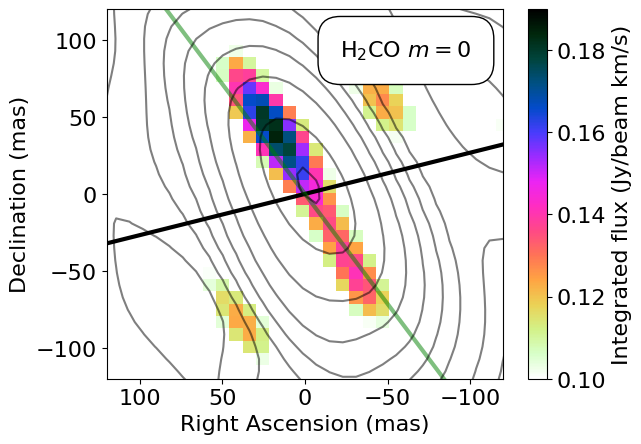}
    \caption{} 
    \label{fig:H2CO_0_zoom}
    \end{subfigure}
\hfill
    \begin{subfigure}[b]{0.45\textwidth}
    \includegraphics[width=\textwidth]{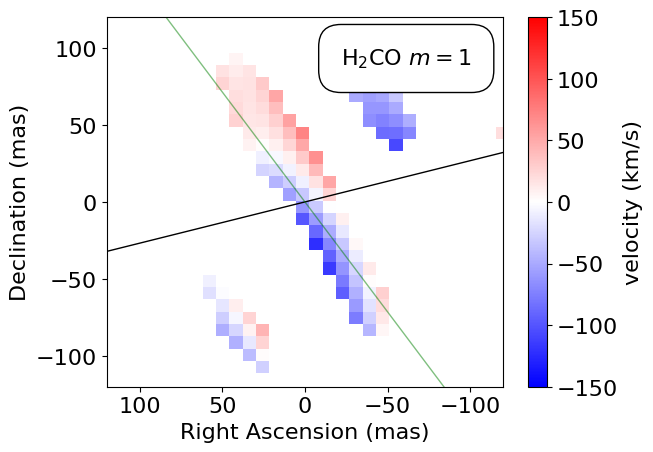}
    \caption{} 
    \label{fig:H2CO_1_zoom}
    \end{subfigure}
\caption{e-MERLIN observations of H$_2$CO $4.83$ GHz emission towards the nuclear region of Zw049.057. (a) moment 0 map of the emission, overlaid with e-MERLIN C Band continuum emission contour map. (b) moment 1 map of the emission. Directions of the CO outflow (PA$_{\mathrm{CO}}=105^{\circ}$), and disk major axis (PA$_{\mathrm{maj}}=30^{\circ}$) are indicated by the black dashed line, black solid line, and green solid line, respectively. Figure is a zoom-in of Fig.~(\ref{fig:H2CO_moment0}) and (\ref{fig:H2CO_moment1}).}
\label{fig:H2CO_zoom}
\end{figure*}

\begin{figure*}[ht!]
\centering
    \begin{subfigure}[b]{0.45\textwidth}
    \includegraphics[width=\textwidth]{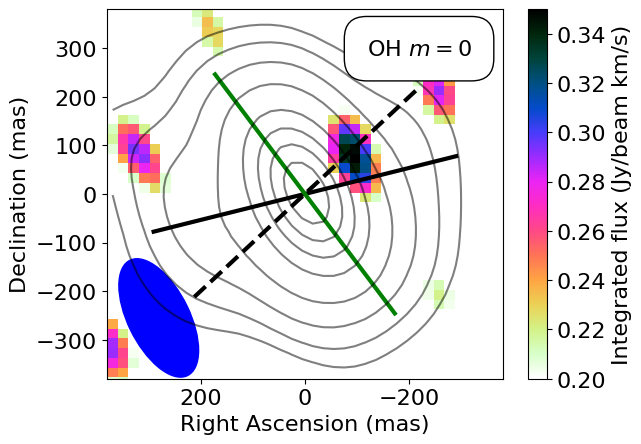}
    \caption{} 
    \label{fig:OH_1665_0}
    \end{subfigure}
\hfill
    \begin{subfigure}[b]{0.45\textwidth}
    \includegraphics[width=\textwidth]{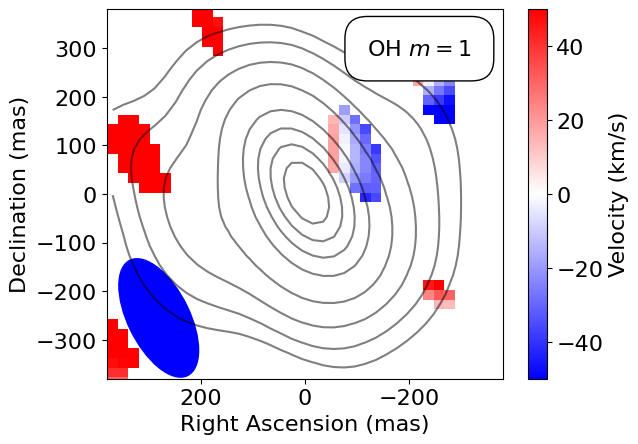}
    \caption{} 
    \label{fig:OH_1665_1}
    \end{subfigure}
\caption{e-MERLIN observations of OH $1.665$ GHz emission towards the nuclear region of Zw049.057. Moment 0 (a) and (1) maps of the emission are overlaid with e-MERLIN L Band continuum emission contour map. Directions of the radio-jet (PA$_{\mathrm{jet}}=130^{\circ}$), CO outflow (PA$_{\mathrm{CO}}=105^{\circ}$), and disk major axis (PA$_{\mathrm{maj}}=30^{\circ}$) are indicated by the black dashed line, black solid line, and green solid line, respectively.}
\label{fig:OH_1665}
\end{figure*}
\end{appendix}

\end{document}